\documentclass[prb,preprint,nofootinbib]{revtex4-1}
\usepackage{amsmath}
\usepackage{amssymb}
\usepackage{graphicx}
\usepackage{url}
\begin{document}
\title{Fermi at Trinity}
\author{J. I. Katz}
\email{katz@wuphys.wustl.edu}
\affiliation{Dept.~Physics and McDonnell Center for the Space Sciences\\
\hbox{Washington University, St.~Louis, Mo. 63130}\\ and\\ \hbox{Los Alamos
National Laboratory, Los Alamos, N.~Mex.~87545}}
\date{\today}
\begin{abstract}
Enrico Fermi estimated the yield of the Trinity test to be about 10 kilotons
by dropping small pieces of paper and observing their motion in the blast
wave.  This is about half the radiochemically derived value of approximately
21 kilotons that necessarily includes thermal and nuclear radiation that do
not contribute to the blast.  Although this story is classic, there appears
to be no account of how he related his observation to the yield.
This note attempts to reconstruct how he might have done so.
\end{abstract}
\maketitle
\section{Introduction}
The official history of Los Alamos\cite{huddeson} and many popular accounts
{describe how Fermi, at Trinity, the first nuclear test, conducted July
16, 1945, estimated its explosive yield with a simple experiment.  He
dropped some small scraps of paper before the blast wave passed by and 
observed how far they were displaced.  From this, he was able to estimate
its explosive yield as 10 kt.  This is about 40\% of the modern estimate
\cite{yield}, but some portion of that was radiated and did not contribute
to the blast wave.  Fermi did remarkably well with a very rudimentary
experiment, but neither he nor subsequent writers appear to have explained
how he inferred the yield.}

\begin{figure*}
	\centering
	\includegraphics[width=\textwidth]{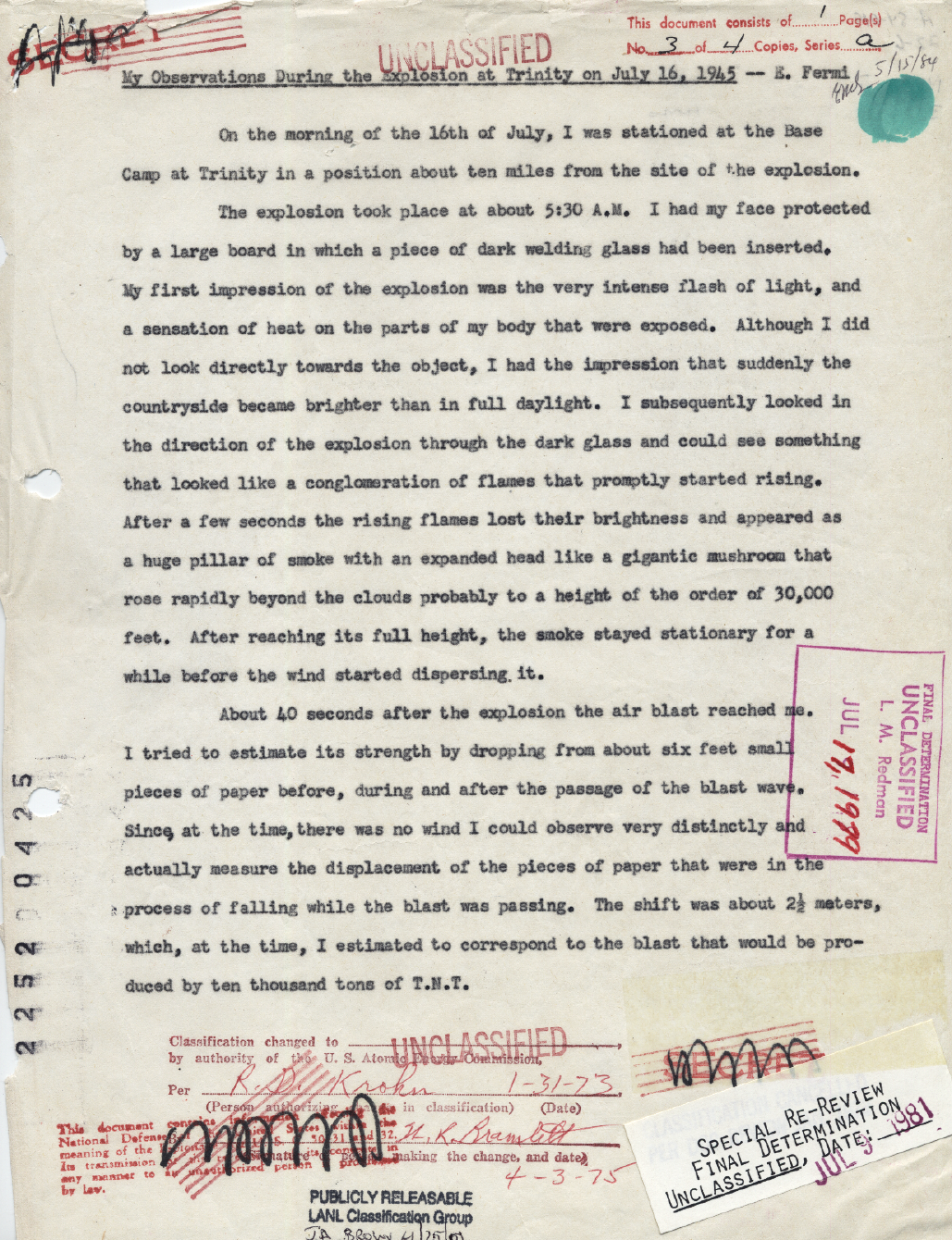}
	\caption{\label{fermiobs} Fermi's original memorandum
	\cite{archive}.}
\end{figure*}

The light scraps of paper that Fermi dropped moved with the air because of
their low ballistic coefficients (the ratio of mass to the product of drag
coefficient and area projected perpendicular to the velocity through the
fluid); {they traced the motion of the air in the blast wave}.  The text
of Fermi's memorandum may be found on a number of web sites, and it is
reproduced in Fig.~\ref{fermiobs}.  This note attempts to reconstruct {his
reasoning}.

{Fermi measured the positive phase air displacement $D$.  In the limit
of a weak blast wave, $D$ is directly related to the positive phase impulse
per unit area $\cal I$, and measuring $D$ may be the best way of measuring
$\cal I$:
\begin{equation}
	D = \int_{v > 0}\!v\,dt = \int_{v>0}\!{P - P_0 \over \rho c_0}\,dt
	= {{\cal I} \over \rho c_0},
\end{equation}
where $\rho c_0$ is the acoustic impedance of the air.}

What was the understanding of blast wave physics at the time of Trinity?
{The theory was developed in some detail in the 1944 Los Alamos document
LA-165\cite{LA165}, and fitted to empirical data on small (mostly 67 lb)
charges in LA-316\cite{LA316} (June 25, 1945, three weeks before Trinity).}
Fermi's responsibilities as a nuclear physicist and experimentalist were
elsewhere, but he was also a leading theorist and likely familiar with these
results.  {However, the results of \cite{LA316} do not extend to
overpressures below 1 lb/in$^2$, approximately ten times those at Fermi's
observation point, and its results for impulse (directly related to the
displacement Fermi measured) for yields $\sim 10,000$ tons relevant to
Trinity do not extend to ranges $> 2$ km, in contrast to the 16 km range to
Fermi's observation, and its results for low impulses are obtained for low
yields but at the unscaled 100 foot height of burst.  Scaling from the May 
7, 1945 100 ton high explosive calibration and rehearsal shot \cite{LA6300}
might have been possible, but no quantitative data from it are documented
in \cite{LA6300}; its purpose appears to have been to test experimental
procedure.  In addition, the differing specific energies of nuclear and
conventional explosives would have made scaling uncertain.}

LA-2000\cite{LA2000}, dated August 1947, two years after Trinity, presents a
comprehensive review edited, and to a substantial extent written, by Hans
Bethe, {extending the earlier theory and presents the results of
numerical calculations apparently performed between 1944 and 1947.}
\cite{LA2000} contains not only sophisticated analytic theory, as befits its
authors, but also in Chap.~6 the results of numerical calculations (``IBM
solution, results or run'') carried out on the primitive computing machines
of the time\cite{Archer}.  Some of those results are used here.

Most of the energy of an airburst is divided into three parts: radiation, 
``wasted'' thermal energy in heated air (partly in the initial fireball and
partly in air irreversibly heated by the blast wave) and an outgoing blast
wave.  A small fraction of the energy appears as radioactive decay and
neutron capture {gamma-rays}.  The displacement of the air (and pieces
of paper that act as Lagrangian tracers of its motion) far from the
explosion includes a net outward displacement that accommodates the
increased volume of a bubble of hot gas (whose subsequent buoyant rise
creates the famous ``mushroom cloud''), and the oscillatory motion of the
outgoing blast wave.
\section{The Bubble}
After pressure equilibrium is achieved, the injection of an energy $Y_b$
into the air increases its volume by
\begin{equation}
	\Delta V = {Y_b (\gamma - 1) \over P_0},
\end{equation}
where $\gamma$ is the adiabatic exponent of air and $P_0$ its initial
pressure.  At temperatures of less than a few thousand K the vibrational
modes of nitrogen and oxygen molecules are not significantly excited and
$\gamma \approx 7/5$.  At higher temperatures $\gamma$ is less; full
classical excitation of the vibrational modes reduces it to $4/3$, and
dissociation and endothermic creation of nitric oxide reduce it still
further.  Most of the volume expansion is attributable to air at lower
temperatures, so here we adopt $\gamma = 7/5$.

For an airburst at the top of a 30 m tower\cite{huddeson} and an observer at
a distance $\gg 30$ m, it is possible to approximate the geometry as
hemispherical.  The effective radius of the hot bubble, defined as the
radius of a hemisphere of volume $\Delta V$,
\begin{equation}
	\label{r0}
	R_H = \left({3 \over 2\pi}{Y_b (\gamma - 1) \over P_0}\right)^{1/3}
	\approx \left({Y_b \over \text{10 kt}}\right)^{1/3} \times 450
	\,\text{m},
\end{equation}
where at the altitude of Trinity of 1.5 km $P_0 \approx 8.5 \times 10^5$
dyne/cm$^2$.  An observer at distance $r$ would see, after all the dynamic
motions have settled down but the hot bubble has not yet begun to rise, an
outward displacement
\begin{equation}
	\Delta r = {\Delta V \over 2 \pi r^2} = {Y_b \over P_0}
	{(\gamma - 1) \over 2 \pi r^2} = {R_H^3 \over 3r^2}.
\end{equation}

For Fermi's observation at $r = 10$ miles (16 km) 
\begin{equation}
	\Delta r \approx \left({Y_b \over 10\,\text{kt}}\right)
	\times 10\,\text{cm},
\end{equation}
where $1\text{ kt (kiloton)} = 4.19 \times 10^{19}$ ergs.  For modern
estimates\cite{yield} that Trinity had a yield of $25 \pm 2$ kt this {gives
25 cm, which} is much
smaller than the 2 1/2 meter displacement he observed, even were there no
radiation losses.  The explanation of Fermi's observation, and the
reconstruction of his argument, must be sought elsewhere.
\section{The Blast Wave}
The sudden creation of a bubble of hot gas produces an outgoing blast wave.
This blast wave is initially very strong, with overpressure $\Delta P \equiv
P - P_0 \gg P_0$, where $P_0$ is the ambient pressure and $P$ the pressure
of the shocked and subsequently rarefied air.  In the strong shock and point
source limit it is described by an analytic theory developed by von
Neumann\cite{vN}, by G.~I.~Taylor\cite{D} and by Sedov, and described {in
this issue} by Baty and Ramsey\cite{BR}.  This is not applicable to the
weak shock regime in which Fermi made his measurement.

\begin{figure}
	\centering
	\includegraphics[width=\columnwidth]{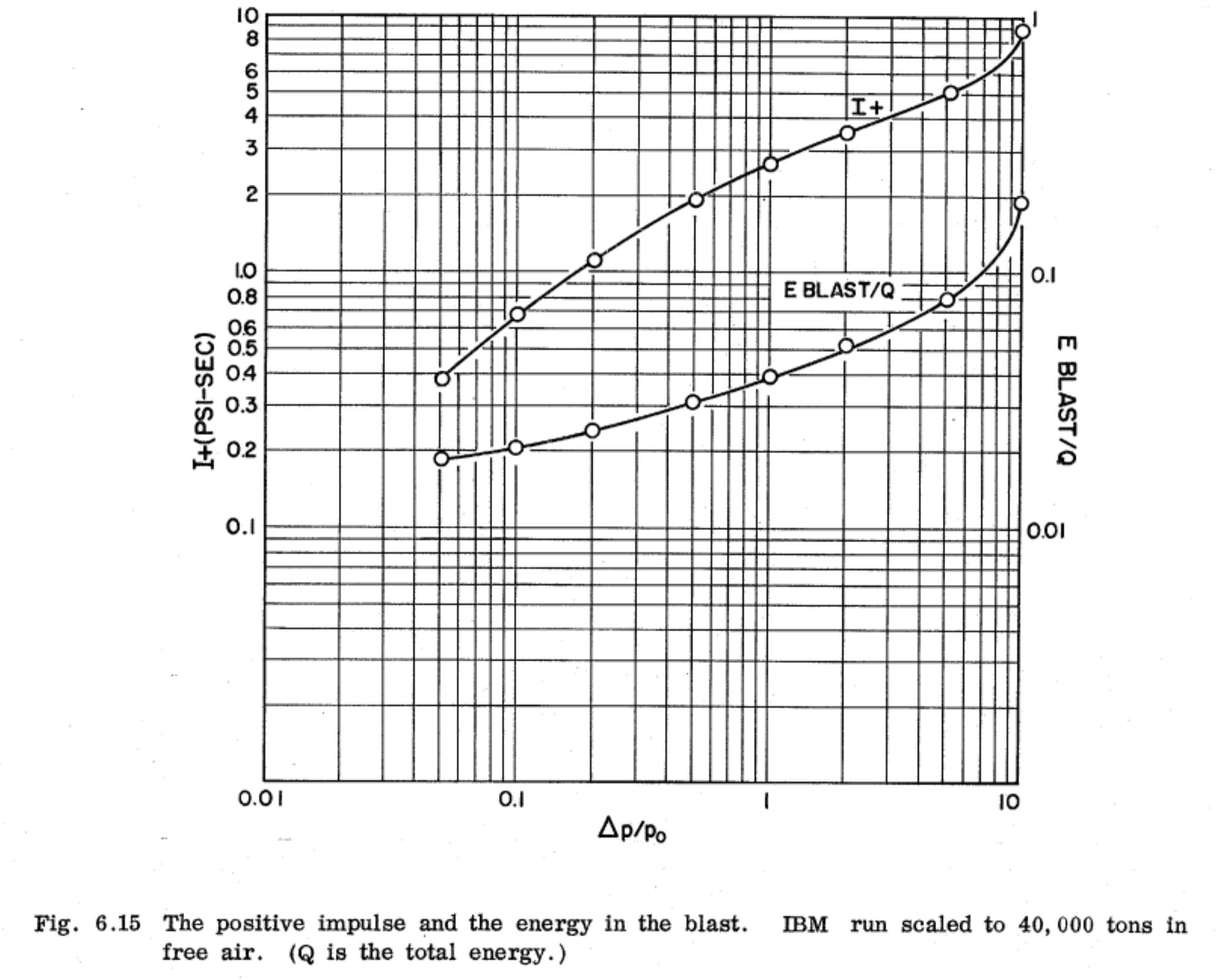}
	\caption{\label{f6.15} Impulse of the positive phase of the blast
	and the fraction of the bomb energy $Q$ that remains in the blast
	wave.  The impulse is scaled to a 40 kt explosion, but the function
	$E_\text{Blast}/Q\,(\Delta P/P_0)$ relating two dimensionless
	parameters is approximately independent of the size of the explosion
	(this is only approximate because scaling breaks down for the
	radiated fraction of $Q)$.  (Fig. 6.15 of LA-2000)}
\end{figure}

LA-2000\cite{LA2000} also calculates weaker blast waves.  However, one
crucial element, the fraction of the explosive yield that appears in the
weak blast wave ($|\Delta P| \ll P_0$) at large distances, where Fermi
measured it, can only be calculated numerically because it involves the
coupling of energy through an intermediate strength regime in which no
analytic theory is valid.  This calculation is described in Chapter 6 of
LA-2000\cite{LA2000}, but it is difficult to reconcile the results there
(Fig.~\ref{f6.15}) with Fermi's measurement and the known yield\cite{yield}
of Trinity.

\begin{figure}
	\centering
	\includegraphics[width=\columnwidth]{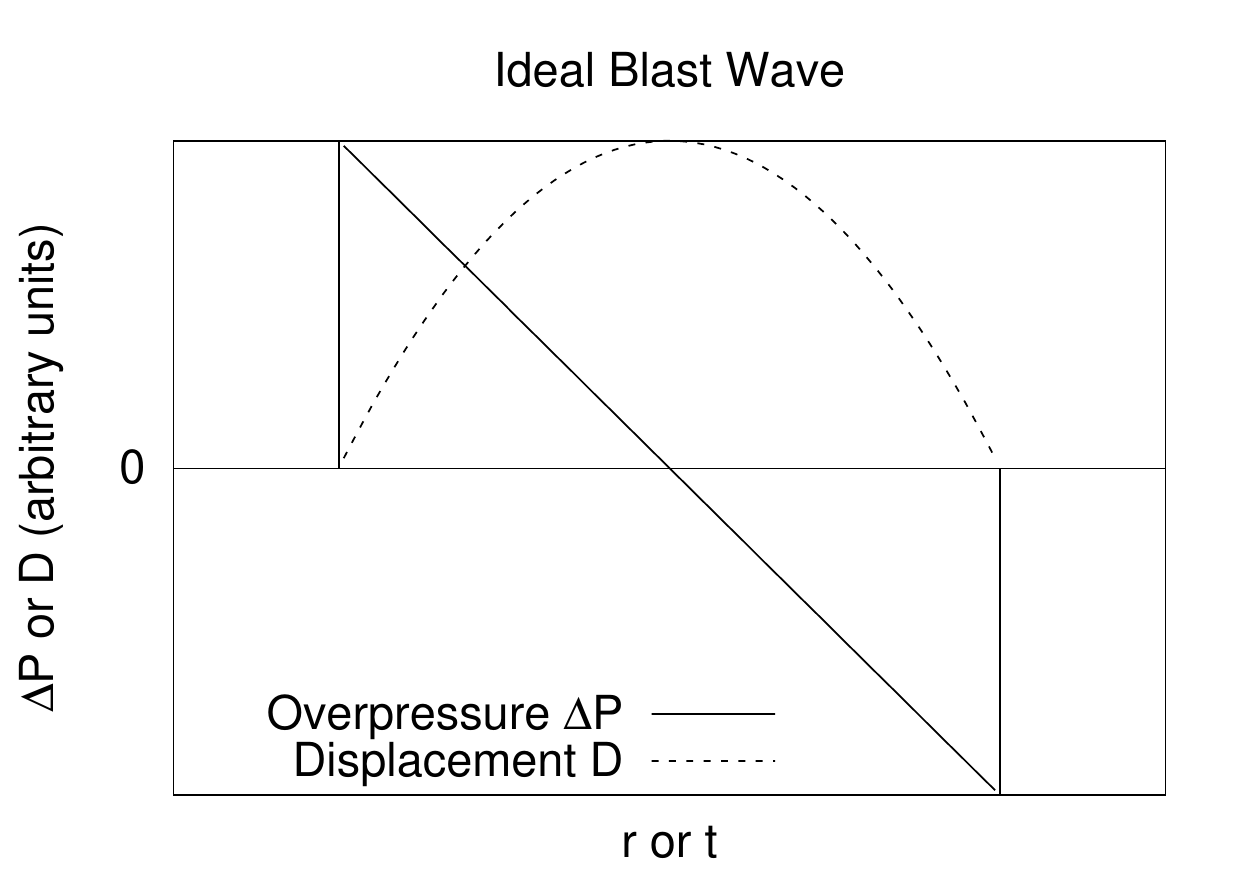}
	\caption{\label{f5.1} Idealized blast overpressure and displacement
	far from the explosion as a function of time at one point or of
	radius at one instant (if plotted as a function of radius the blast
	is propagating to the left and the location of the explosion
	asymptotically distant to the right).  This profile is the classic
	``N-wave''\cite{CF48}, named for its shape; it is produced when an
	idealized membrane between two regions of infinitesimally differing
	pressures is abruptly removed.  This pressure profile is shown in 
	Fig.~5.1 of LA-2000.  The displacement profile is added here, and
	both were familiar to Fermi.  The rarefaction phase of the blast
	produced by a strong explosion is quite different because there are
	no rarefaction shocks of finite amplitude; Fig.~\ref{f6.13} shows
	the results of an early calculation from LA-2000.  Far from the
	explosion the blast propagates as an acoustic wave with nearly
	constant profile, but with amplitude decreasing in proportion to
	$1/r$.}
\end{figure}

Once the blast wave becomes weak, it is, to a good
approximation, a sound wave propagating at the sound speed.  The over- and
under-pressure $\Delta P$ and the fluid speed decay nearly proportionally to
$1/r$ in spherical or hemispherical geometry.  The width of the blast wave
is nearly constant because the sound speed is nearly its ambient value
throughout.  Idealized displacement and overpressure profiles are shown in
Fig.~\ref{f5.1}.  Unlike the bubble displacement, in the acoustic limit the
blast wave displacement returns to zero after the wave passes.

\begin{figure}
	\centering
	\includegraphics[width=\columnwidth]{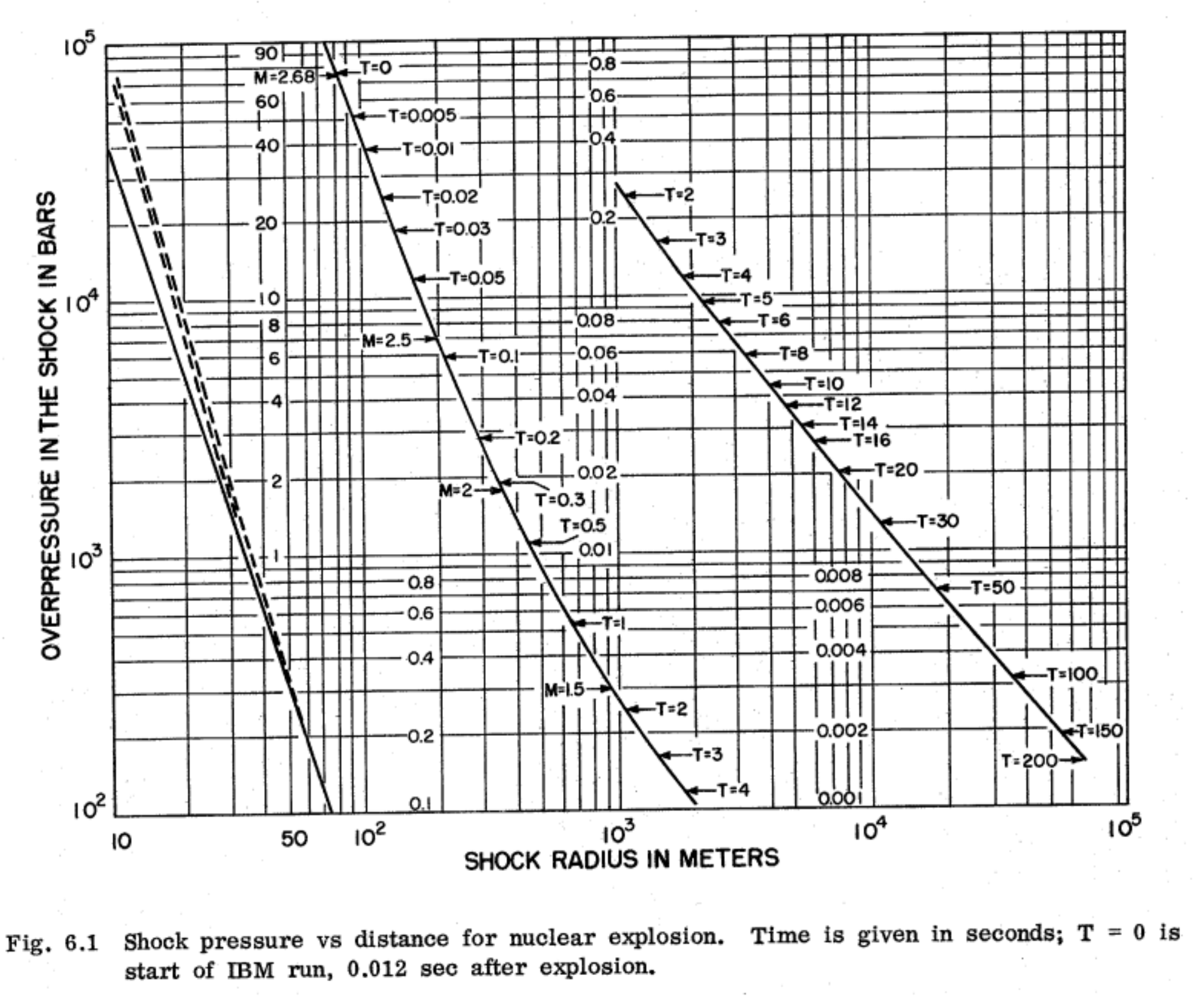}
	\caption{\label{f6.1} Calculated shock overpressure {\it vs.\/}
	radius for a 13 kt explosion in ambient air in spherical geometry,
	corresponding to a 6.5 kt explosion in hemispherical geometry.  Note
	the separate ordinate labels, arrival times $T$ and Mach numbers $M$
	for each curve.  At Fermi's distance of 16 km, in hemispherical
	geometry the overpressure would have been 0.008 bar for a 6.5 kt
	explosion and 0.010 bar for 13 kt.  $\Delta P$ scales approximately
	$\propto Q^{1/3}$, where $Q$ is the yield, because the energy
	density is $\propto (\Delta P)^2$ in an acoustic wave and the blast
	wave thickness scales $\propto R_H \propto Q^{1/3}$.  (Fig.~6.1 of
	LA-2000)}
\end{figure}

The dependence of shock overpressure on distance from the explosion is shown
in Fig.~\ref{f6.1} for a spherically symmetric explosion into air at ambient
conditions ($P_0 = 1.01 \times 10^6\,$dyne/cm$^2$ and $T = 300\,$K).  This
Figure combines the numerical (``IBM'') results with the ``semi-acoustic''
and acoustic theory for weak and very weak (acoustic) blast waves.  \S 6.3
of LA-2000 states that the calculation was intended to describe a 10 kt
explosion but actually described a 13 kt (not including thermal or nuclear
radiation) explosion.

\begin{figure}
	\centering
	\includegraphics[width=\columnwidth]{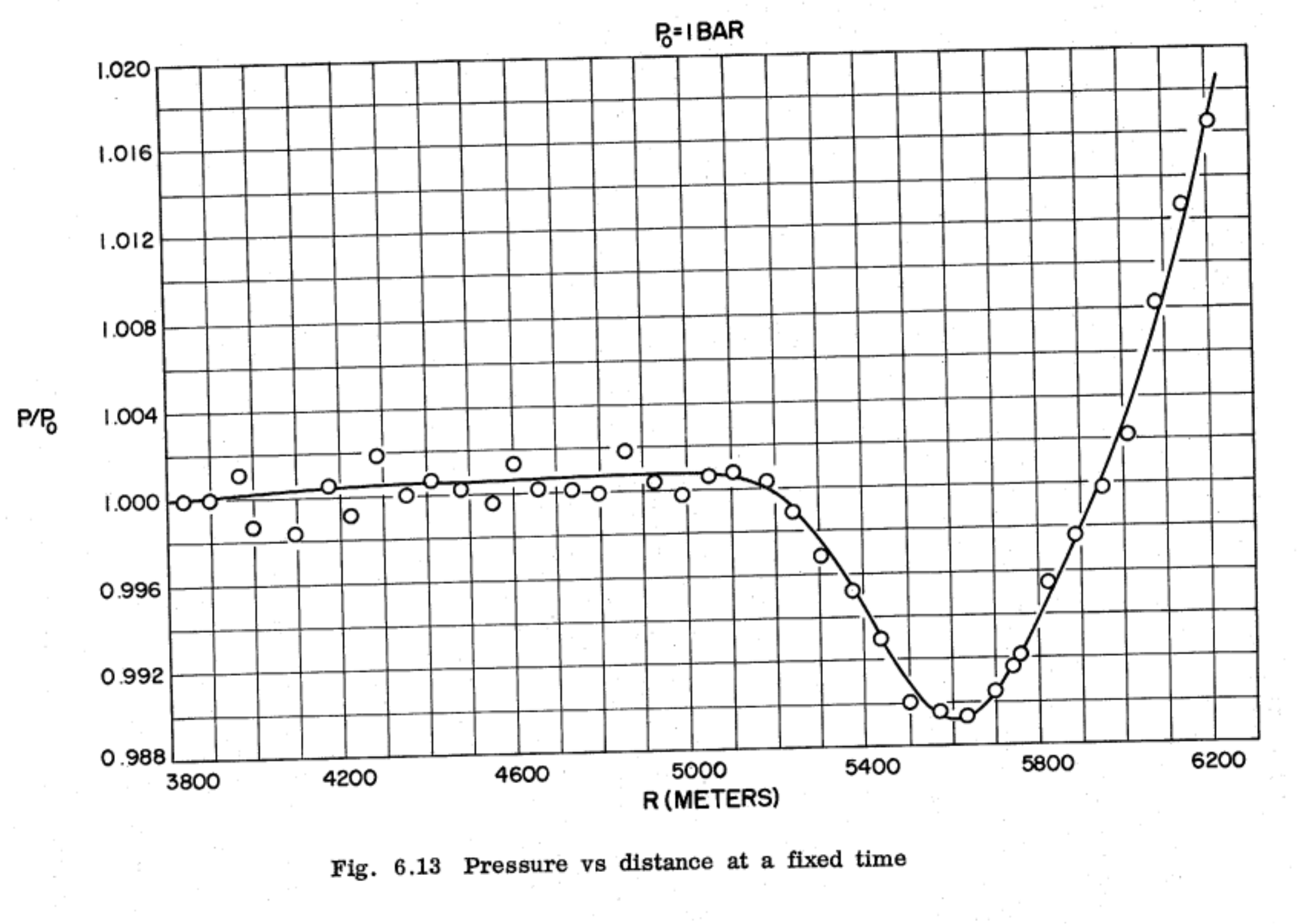}
	\caption{\label{f6.13} Calculated blast wave pressure profile far
	from the explosion for a yield of 13 kt in spherical geometry or 6.5
	kt in hemispherical geometry.  The width and profile $P(R)$ of the
	blast wave vary only very slowly from this radius (the greatest
	radius for which this is shown in LA-2000) outward, while the
	overpressure $\Delta P$ scales $\propto 1/r$.  This profile is not
	close to the ideal profile of Fig.~\ref{f5.1}, indicating that the
	model is only approximate.  At the shock $P/P_0$ abruptly rises from
	unity, but the rise (a vertical line in Fig.~\ref{f5.1}) is not
	shown here.  According to \S 6.4 of LA-2000, $\Delta P = 0.0251 P_0$
	at 6270 m with a positive phase 290 m wide, but the portion of the
	pressure curve with $\Delta P > 0.016 P_0$ and $r > 6200\,$m is not
	shown.  If the shock is actually at 6270 m, as stated, the positive
	phase is 320 m wide.  The negative phase is much wider.  (Fig.~6.13
	of LA-2000)}
\end{figure}

\begin{figure}
	\centering
	\includegraphics[width=\columnwidth]{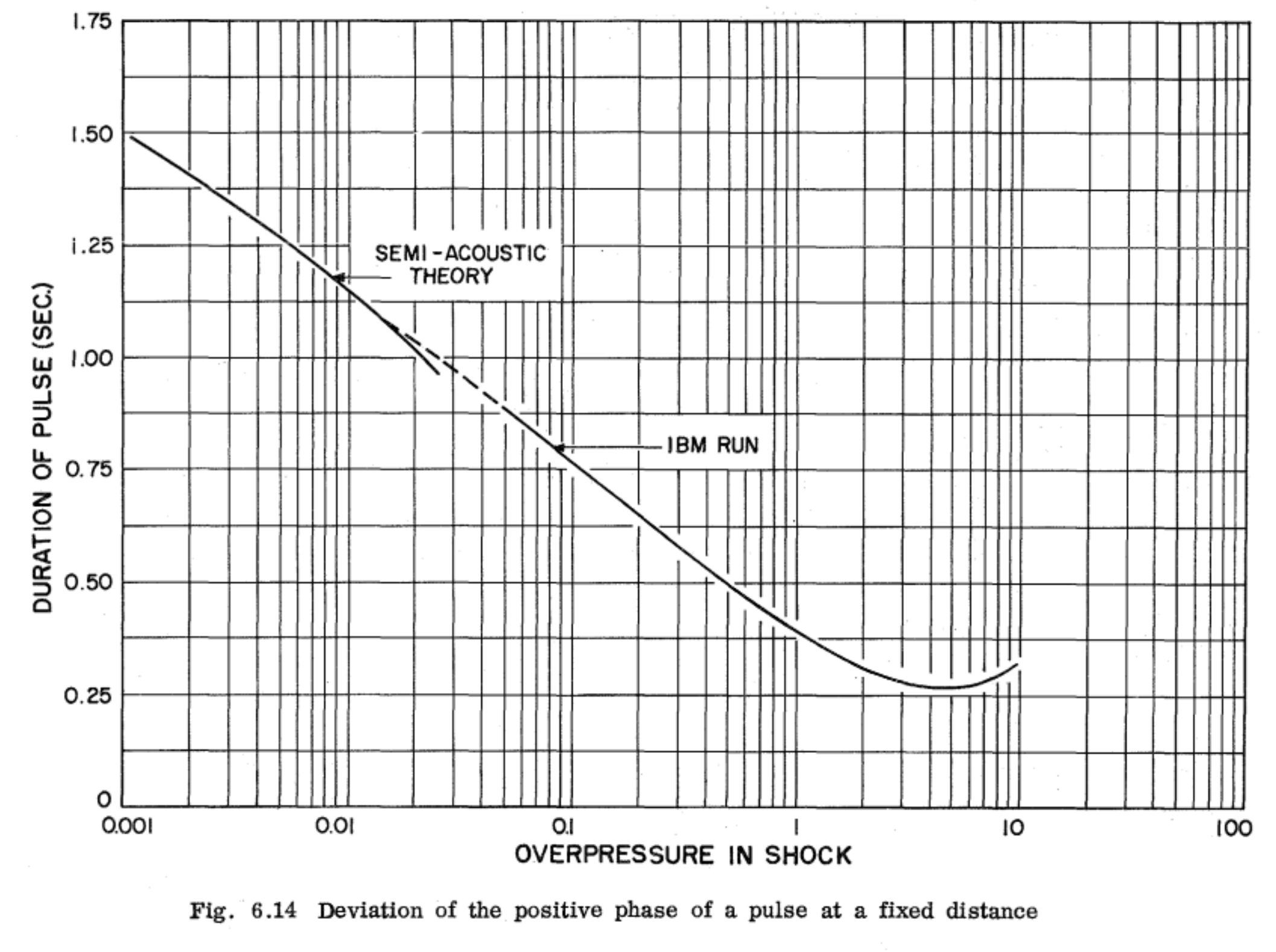}
	\caption{\label{f6.14} Duration $\Delta t$ of the positive phase as
	a function of overpressure (original caption has ``Deviation'' for
	``Duration'').  The blast wave width increases logarithmically even
	in the acoustic regime because the overpressure, although small, is
	not infinitesimal.  The ``IBM Run'' was the full numerical
	calculation in the moderately strong and strong shock regimes, while
	the ``semi-acoustic'' theory of LA-2000 calculates the flow to
	second order in $\Delta P/P_0$; a first-order calculation is
	strictly valid only in the acoustic limit.  In this limit the
	positive phase duration $\Delta t$ and thickness $h$ are related by
	$h = c_0 \Delta t$.  (Fig.~6.14 of LA-2000)}
\end{figure}

For $r \gg R_H$ the leading edge of the blast wave becomes a weak shock
($\Delta P \ll P_0$) that performs very little irreversible work
($\propto (\Delta P)^3$).  The blast wave propagates at nearly the speed of
sound, $c_0 = 340\,$m/s in air at $15^{\,\circ}$C, a typical temperature at
dawn in July at Trinity (LA-2000 takes $c_0 = 347\,$m/s, appropriate to air
at $27^{\,\circ}$C).  The passage of a weak shock produces a velocity
discontinuity
\begin{equation}
	\label{imped}
	\Delta v = {\Delta P \over \rho c_0},
\end{equation}
defining the acoustic impedance $\rho c_0$, where $\rho$ is the ambient
density (very close to the shocked density).  The pressure profile of the
weak blast wave at one radius is shown in Fig.~\ref{f6.13}, and the duration
of the positive phase ($P > P_0$) is shown in Fig.~\ref{f6.14}, both for a
yield of 13 kt in spherical symmetry.

In this weak blast wave regime the rate of dissipation of blast energy by
conversion to thermal ``waste energy'', so-called because it does not
contribute to the destructive effect of the blast, rapidly decreases with
increasing $r$.  However, it decreases slowly enough that the energy in the
blast wave does not approach a constant asymptote, but rather declines
logarithmically with overpressure and distance (Eq.~429 of
LA-165\cite{LA165}).

The energy of a weak hemispherical blast wave whose positive phase has
thickness $h$
\begin{equation}
	\label{EB}
	E_\text{Blast} = 2 \pi r^2 h \times 2 \times 2 \times {1 \over 3}
	\times {\rho (\Delta v)^2 \over 2}.
\end{equation}
One factor of 2 results from the contribution of the compression energy,
equal to the kinetic energy in an acoustic wave, and another factor of 2
from the contribution of the negative phase, nominally equal to that of
the positive phase (Fig.~\ref{f5.1}, although Fig.~\ref{f6.13} suggests a
factor of 1.5 may be more accurate), and the factor of $1/3$ comes from
averaging $v^2$ with $v$ varying linearly from $0$ to $\Delta v$ through the
blast wave (Fig.~\ref{f5.1}).  The thickness $h$ is comparable to the radius
$R_H$ where the shock transitions from strong to weak, scaling $\propto
Q^{1/3}$, but grows logarithmically (LA-165\cite{LA165} and
Fig.~\ref{f6.14}) as the weak blast wave propagates.

Fermi had to estimate $h = c_0 \Delta t$ theoretically because it would not
have been easy to measure $\Delta t$ without photographic or electronic
equipment.  At $\Delta P = 0.008 P_0$ (16 km from a 13 kt explosion in
spherical geometry; Fig.~\ref{f6.1}) the duration of the positive phase
$\Delta t = h/c_0 = 1.14\,$s for the same explosion (Fig.~\ref{f6.14}).
Scaling to 13 kt in hemispherical geometry while holding $\Delta P$ constant
doubles the effective yield and multiplies $R_H$ and $\Delta t$ by
$2^{1/3}$.  Then $\Delta t = 1.44\,$s and $h = c_0 \Delta t \approx 500\,$m.
This is (probably fortuitously) very close to the $R_H = 490\,$m estimated
from Eq.~\ref{r0}.

At $\Delta P = 0.02P_0$, as shown in Fig.~\ref{f6.13} at 6 km for 13 kt in
spherical geometry, Fig.~\ref{f6.14} indicates $\Delta t = 1.04\,$s and $h =
360\,$m rather than the 320\,m indicated by Fig.~\ref{f6.13} or the 290 m of
\S 6.4 of LA-2000.  This should be taken as a measure of the resolution and
precision of the calculations.

From the results reported in LA-2000 (or the more approximate elementary
analytic result Eq.~\ref{r0}) Fermi had a good idea of the value of $h$ at
his observation point for yields in the expected range.  Fortunately,
$\Delta t$ and $h$ depend only logarithmically on $\Delta P$
(LA-165\cite{LA165} and Fig.~\ref{f6.14}).  In principle, he could have used
the computed dependence of $h$ on yield to solve interatively for the yield,
but this dependence is so weak that such a procedure was unnecessary.
Measuring a displacement $D = 2.5\,$m provided the peak velocity $\Delta v$
for the linear pressure and velocity profile of Fig.~\ref{f5.1}.  Then
\begin{equation}
	\label{v}
	\Delta v = {2 D c_0 \over h} = 3.4\ \text{m/s}.
\end{equation}
This is fairly close to the $v = 2.9\,$m/s calculated for $\Delta P =
0.010\,$bar (scaling the results of Fig.~\ref{f6.1} to hemispherical
geometry by multiplying $\Delta P$ by $2^{1/3}$) from Eq.~\ref{imped},
taking $\rho = 1.0\,$kg/m$^3$ at the 1500 m altitude of Trinity and noting
that $\Delta P$ scales $\propto E_\text{Blast}/r^3$, independent of the air
density.  Substituting $h$ and $\Delta v$ in Eq.~\ref{EB} yields 
\begin{equation}
	\label{eblast}
	E_\text{Blast} = 1.5\,\text{kt}.
\end{equation}

In order to relate his observation to the explosive yield Fermi had to know
what fraction of the yield appears as blast wave energy at an overpressure
of about 0.01 bar.  If he had relied on the IBM calculation reported in
Chap. 6 of LA-2000, shown in Fig.~\ref{f6.15}, he would have been seriously
misled.  A small extrapolation from these results to $\Delta P = 0.01\,$bar
would have led to $E_\text{Blast} \approx 0.016 Q$, implying an impossible
$Q \approx 90\,$kt, far from both Fermi's estimate and the modern
radiochemical value of $25 \pm 2$ kt\cite{yield} for the total yield, that
includes thermal and nuclear radiation.

Two factors may contribute to this discrepancy.   The effects of numerical
viscosity were likely not appreciated when the IBM calculations were done,
and the decline of $E_\text{Blast}/Q$ with increasing range in
Fig.~\ref{f6.15} may be mostly a numerical artefact.  In addition, the
``wasted'' energy contributes to the pressure driving the blast wave, so it
is not really wasted.  These hypotheses could be readily tested with a
modern calculation.

How did Fermi get it right?  He could have readily estimated $E_{Blast}$
from Eqs.~\ref{EB} and \ref{v} or equivalent forms.  He would have needed a
value for $h$, but that could have been taken from the numerical
calculations (were they available and known to him in 1945), estimated in
advance of Trinity as the shock radius $R_H$ (Eq.~\ref{r0}) where $\Delta P
= {\cal O}(P)$, or inferred at Trinity from eyeball observation of the
duration of the positive phase using his paper Lagrangian tracers.  He
didn't tell us what he did.

Fermi would also have needed a value for $E_\text{Blast}/Q$, an essential
parameter that the only possibly available numerical calculation got wrong.
{He was likely familiar with the Bethe-Kirkwood-Penney result (Eq.~429
of LA-165\cite{LA165})
\begin{equation}
	\label{loglaw}
	{E_{Blast} \over Q} \approx {1 \over 3 \sqrt{\ln{P_0/\Delta P}}}
	\approx 0.15,
\end{equation}
which, combined with his empirical value for $E_{Blast}$ (Eq.~\ref{eblast})
immediately leads to his yield estimate $Q \approx 10$ kt.}
%

If the IBM calculations had been done by the time of Trinity, either Fermi
was unaware of them or wisely chose to use the analytic theory rather than
those apparently quantitative computational results.
\section{Conclusion}
{At the time Fermi wrote his famous memorandum, the result
(Eq.~\ref{loglaw}) of LA-165\cite{LA165} was likely well-known at Los Alamos,
so that no explanation was necessary.  This theory is no longer common
knowledge, even though the original documents are readily available to
anyone.}

Fermi estimated the yield of Trinity to be 10 kt, about 40\% of the modern
value, from the motion of Lagrangian tracers, his famous scraps of paper.
Significant additional energy was radiated, but how much depends on the
physics and chemistry of hot air (in particular, the thermodynamics and
opacity of oxides of nitrogen) and requires numerical methods to calculate.
Fermi's result was therefore a lower bound, and was sufficient to declare
Trinity a success.
\section{Acknowledgments}
I thank E.~Deschamp of the Los Alamos Reports Library for a scan of Fermi's
memorandum (Fig.~\ref{fermiobs}), B.~Albright, S.~Andrews, M.~B.~Chadwick
and R.~L.~Garwin for discussions, and an anonymous referee for calling my
attention to the pre-Trinity Los Alamos reports.

This work was supported by the US Department of Energy through the Los
Alamos National Laboratory. Los Alamos National Laboratory is operated
by Triad National Security, LLC, for the National Nuclear Security
Administration of the US Department of Energy under Contract No.
89233218CNA000001.


\end{document}